\documentclass[10pt]{article}
\usepackage{amssymb,bbm,amsthm}
\usepackage{amsmath}
\usepackage{amsbsy}
\usepackage{cite}
\usepackage[dvips]{graphicx}
\newcommand{\as}{\\[.6em]}

\newcommand{\AS}{\\[1.2em]}
\newcommand{\dis}{\displaystyle}
\newcommand{\bela}[1]{\begin{equation}\label{#1}}
\newcommand{\ela}{\end{equation}}
\newcommand{\bear}[1]{\begin{array}{#1}}
\newcommand{\ear}{\end{array}}

\newcommand{\bx}{\mbox{\boldmath $x$}}

\newcommand{\bu}{\mbox{\boldmath $u$}}

\newcommand{\bt}{\mbox{\boldmath $t$}}

\newcommand{\sfK}{\mathsf{K}}

\newcommand{\Z}{\mathbbm{Z}}

\newcommand{\R}{\mathbbm{R}}

\newcommand{\n}{\mbox{\boldmath $n$}}

\newcommand{\phib}{\mbox{\boldmath $\phi$}}
\newcommand{\Ib}{\mbox{\boldmath $I$}}

\newcommand{\Dbb}{\mathbbm{D}}

\theoremstyle{plain}

\theoremstyle{definition}

\numberwithin{equation}{section}

\begin{document}

\begin{center} 
{\Large\bf Integrable discretisation of hodograph-type systems, hyperelliptic integrals\vspace{2mm} and\\ Whitham equations}
%{\Large\bf Discretisation of Whitham and \vspace{2mm} hodograph-type equations and associated %hyperelliptic integrals}
 %{\Large\bf Generalised hodograph equations, systems of\vspace{2mm}  hydrodynamic type and their %discretisation}
\end{center}

\smallskip

\begin{center}
 \sc
 B.G.\ Konopelchenko$\,^{1}$ and W.K.\ Schief$\,^{2,3}$
\end{center}

\smallskip

\begin{center}
 \small\sl
$^1$ Department of Mathematics and Physics ``Ennio de Giorgi'', University of Salento and sezione INFN, Lecce, 73100, Italy\\[2mm]
$^2$ School of Mathematics and Statistics, The University of New South Wales, Sydney, NSW 2052, Australia\\[2mm]
$^3$ Australian Research Council Centre of Excellence for Mathematics and Statistics of Complex Systems, School of Mathematics and Statistics, The University of New South Wales, Sydney, NSW 2052, Australia
\end{center}

\begin{abstract}
Based on the well-established theory of discrete conjugate nets in discrete differential geometry, we propose and examine discrete analogues of important objects and notions in the theory of semi-Hamiltonian systems of hydrodynamic type. In particular, we present discrete counterparts of (generalised) hodograph equations, hyperelliptic integrals and associated cycles, characteristic speeds of Whitham type and (implicitly) the corresponding Whitham equations. By construction, the intimate relationship with integrable system theory is maintained in the discrete setting.
\end{abstract}

\section{Introduction}

Systems of quasi-linear first-order differential equations of the form
\bela{B1}
  u_{t}^{i}=\lambda ^{i}(\bu)u_{x}^{i},\qquad i=1,\ldots,N + 1,
\ela
where subscripts denote partial derivatives, represent an
important subclass of partial differential equations which admit special properties and a variety of
applications \cite{1}. In physics, such systems arise, in particular, as limits of nonlinear partial differential equations without dissipation or dispersion and as Whitham equations for slow modulations (see, e.g., \cite{2,3}). The theory of (more general) Hamiltonian quasi-linear systems of hydrodynamic type has been developed by Dubrovin and Novikov \cite{4,5,6}. It has been established in \cite{7,8,9,Per70,FisPer76,Gru84} that these systems are intimately related
to important notions in classical differential geometry. In particular, it has been demonstrated by Tsarev \cite{7,8,9} that a semi-Hamiltonian system of the type (\ref{B1}) possesses an infinite set of integrals of motion with densities $\psi$ obeying the system of linear hyperbolic equations 
\bela{B2}
  \psi_{u^{i}u^{k}}=A_{ik}\psi_{u^{i}}+A_{ki}\psi_{u^{k}},\qquad i\neq k,
\ela
wherein the coefficients $A_{ik}$ are defined by the system
\bela{B4}
  \lambda^i_{u^k} = A_{ik}(\lambda^k - \lambda^i),
\ela
and the fluxes $\psi_\star$ in the corresponding conservation laws are (uniquely) determined via integration of the compatible system
\bela{B4a}
  \psi_{\star u^i} = \lambda^i\psi_{u^i}.
\ela
These semi-Hamiltonian systems admit an infinite number of symmetries \cite{7,8,9}
\bela{B3}
  u_{t^{\alpha }}^{i}=\lambda _{\alpha }^{i}(\bu)u_{x}^{i},\qquad\alpha =2,3,4, \ldots,
\ela
where each set of characteristic speeds $\{\lambda^i_\alpha\}$ constitutes a solution of (\ref{B4}) regarded as a linear system. The compatibility of the latter (in the sense of a natural Cauchy problem) is equivalent to the existence of the semi-Hamiltonian structure. In the differential-geometric context, the constituent equations of (\ref{B2}) are known as conjugate net equations since these constitute the governing linear equations in the classical theory of conjugate nets (see, e.g., \cite{10}). Moreover, the connection between the density $\psi$ and the flux $\psi_\star$ is of Combescure type (see, e.g., \cite{RogersSchief2002}).  In modern integrable system terminology, the densities $\psi$ constitute eigenfunctions while the characteristic speeds $\lambda^i_\alpha$ represent associated adjoint eigenfunctions.

Remarkably, Tsarev has proven \cite{7,8} that, locally, all solutions of a semi-Hamiltonian system of the form (\ref{B1}) are given implicitly by the algebraic system
\bela{B5}
  x + \lambda ^{i}(\bu)t -\omega^{i}(\bu)=0,\qquad i=1,\ldots,N+1
\ela
with $\{\omega^i\}$ denoting the general set of adjoint eigenfunctions obeying the linear system (\ref{B4}). This linearisation technique has come to be known as the generalised hodograph method since, in the case $N=1$, the quantities $x$ and~$t$ regarded as the unknowns of the system (\ref{B5}) obey the classical hodograph equations \cite{CourantFriedrichs1948}
\bela{B5a}
  x_{u^1} + \lambda^2(\bu) t_{u^1} = 0,\quad x_{u^2} + \lambda^1(\bu) t_{u^2} = 0.
\ela
In fact, in this paper, it is shown that such (generalised) hodograph equations exist for arbitrary $N$. In summary, the properties of semi-Hamiltonian systems of the form (\ref{B1}) and their solutions are completely encoded in the classical surface theory of conjugate nets. This highlights the privileged nature of semi-Hamiltonian systems of hydrodynamic type.

The particular class of conjugate nets governed by the compatible hyperbolic equations 
\bela{B6}
  \phi _{x^{i}x^{k}}=\frac{1}{x^{i}-x^{k}}(\epsilon^{k}\phi _{x^{i}} - \epsilon^{i}\phi _{x^k}),
\ela
wherein the $\epsilon^i$ constitute constants,
plays a distinguished role in the theory of semi-Hamiltonian hydrodynamic-type systems (with the identification $\bu=\bx$). However, it is important to note that, in general, these special conjugate net equations are, {\em a priori}, unrelated to the conjugate net equations (\ref{B2}). In particular, the eigenfunction $\phi$ does not necessarily play the role of a density. The linear equations (\ref{B6}) are known as Euler-Poisson-Darboux equations and have been the subject of extensive investigation in classical differential geometry (see, e.g., \cite{12}). Their importance in the one-phase Whitham equations for the Korteweg-de Vries and nonlinear Schr\"odinger equations has been observed in \cite{13,14,15} and, in the multi-phase case, in \cite{16}. In fact, the explicit expressions for the characteristic speeds in the multi-phase Whitham equations derived in the pioneering paper \cite{17} and
also in \cite{16} contain, as elementary building blocks, particular solutions of the Euler-Poisson-Darboux equations with parameters $\epsilon^i = \frac{1}{2}$. Recently it has been demonstrated \cite{18} that
the characteristic speeds for multi-phase Whitham equations may be obtained by means of iterated Darboux transformations generated by contour integrals of separable solutions of (extended) Euler-Poisson-Darboux systems. Euler-Poisson-Darboux systems for different values of $\epsilon^i$ play also a central role in the treatment of various dispersionless soliton equations and $\epsilon$-systems \mbox{\cite{19,20,21}}.

The connection between the Euler-Poisson-Darboux system and the characteristic speeds of multi-phase Whitham equations (and therefore the associated conjugate net system (\ref{B2})) is provided by the observation that the hyperelliptic integrals
\bela{B7}
  \oint_{b_\kappa}\frac{\zeta^k}{\sqrt{\prod_{i=1}^{2g+1}(\zeta - x^i)}}d\zeta,
\ela
where the contours $b_\kappa$, $\kappa=1,\ldots,g$ are appropriately chosen cycles \cite{16,17,hyper} and $k\leq g-1$, may be regarded as superpositions of separable solutions of the Euler-Poisson-Darboux system (\ref{B6}) for $\epsilon^i=\frac{1}{2}$. The methods recorded in \cite{16,17,18} are then used in the (algebraic) construction of the characteristic speeds $\lambda^i$. For instance, in the case $g=1$, the elliptic integral (\ref{B7}) may be calculated to be essentially
\bela{B8}
  \phi^{\rm O} = \frac{2}{\pi\sqrt{x^3-x^1}}\sfK\left(\sqrt{\frac{x^3-x^2}{x^3-x^1}}\right),
\ela
where $\sfK$ denotes the complete elliptic integral of the first kind. Each of the three coordinates $x^i$ gives rise to a classical Levy transform \cite{Eisenhart1962}
\bela{B9}
  \lambda^i = \phi^0 - \frac{\phi^{\rm O}}{\phi^{\rm O}_{x^i}}\phi^0_{x^i}
\ela
of the simplest non-constant solution $\phi^0 = \frac{1}{2}(x^1 + x^2 + x^3)$ of the Euler-Poisson-Darboux system (\ref{B6}) generated by $\phi^{\rm O}$ and these coincide with the characteristic speeds for the one-phase Whitham equations \cite{2}. The avatar (\ref{B9}) of these characteristic speeds may be found in \cite{15,Kudashev91}. It is remarked in passing that the action of the Levy transformation on semi-Hamiltonian systems of hydrodynamic type has been discussed in detail in \cite{Ferapontov2000}.

As indicated above, many systems of hydrodynamic type (\ref{B1}) admit dispersive counterparts which are integrable by means of the Inverse Spectral Transform (IST) method (see, e.g., \cite{2,6}). One of the
remarkable properties of IST integrable equations is that they admit integrable discretisations which reveal their fundamental properties (see, e.g., \cite{Suris03}). Such discretisations are usually constructed
via invariances of the integrable equations under B\"acklund,
Darboux or similar discrete transformations (see, e.g., \cite{RogersSchief2002}). This method simultaneously leads to a discretisation of the underlying linear representation (Lax pair \cite{AblowitzSegur1981}). 

Based on the standard discretisation (see, e.g., \cite{11}) of the conjugate net equations (\ref{B2}) and associated adjoint equations (\ref{B3}), we here propose a canonical integrability-preserving way of discretising the theory outlined in the preceding. In particular, this is shown to lead to integrable discretisations of generalised hodograph equations, canonical cycles and associated hyperelliptic integrals and characteristic speeds of commuting flows of hydrodynamic type such as those corresponding to the multi-phase Whitham equations. It is noted that large classes of solutions of the standard discrete conjugate net equations may be obtained by means of, for instance, the $\partial$-bar dressing method \cite{KonopelchenkoBogdanov95}, Darboux-type transformations \cite{ManDolSan97,LiuMan98} or the algebro-geometric approach employed in \cite{AkhmetshinKricheverVolvoski99}. Our approach exploits the existence of a canonical discretisation of the classical Euler-Poisson-Darboux system (\ref{B6}) and associated separable solutions.

\section{Generalised hodograph equations}

We are concerned with commuting flows of diagonal systems of hydrodynamic type, that is, compatible systems of first-order equations of the type
\bela{G1}
  u^i_{t^\alpha} = \lambda^i_{\alpha}(\bu)u^i_x,\qquad i = 1,\ldots, N + 1,\quad \alpha = 1,\ldots,N,
\ela
where the subscripts on the functions $u^i$ denote derivatives with respect to the independent variables $x$ and $t^\alpha$. It is emphasised that even though the above systems constitute the point of departure, many of the mathematical notions presented in this paper go beyond these systems and turn out to be of interest in their own right. It is known \cite{8} that diagonal systems of hydrodynamic type commute if and only if the $N$ sets of  characteristic speeds $\{\lambda^i_{\alpha}\}$ labelled by $\alpha$ obey the same linear system
\bela{G3}
  \lambda^i_{\alpha u^k} = A_{ik}(\lambda^k_{\alpha} - \lambda^i_{\alpha}),
\ela
where the coefficients $A_{ik}$ may be regarded as being defined by the equations for, say, $\alpha=1$. Here, $\lambda^i_{\alpha u^k}=\partial\lambda^i_\alpha/\partial u^k$. Furthermore, it is readily verified that the above linear equations may also be regarded as the compatibility conditions for the existence of some functions $\psi_\alpha$ defined (up to constants of integration)~by 
\bela{G4}
  \psi_{\alpha u^i} = \lambda^i_{\alpha}\psi_{u^i},
\ela
where $\psi$ is a solution of the linear hyperbolic equations
\bela{G2}
  \psi_{u^iu^k} = A_{ik}\psi_{u^i} + A_{ki}\psi_{u^k}.
\ela
It is important to note that the coefficients $A_{ik}$ cannot be arbitrary as these are constrained by the compatibility conditions for the hyperbolic equations (\ref{G2}) or, equivalently, the first-order equations (\ref{G3}). In fact, the coefficients $A_{ik}$ must be solutions of an integrable system of nonlinear partial differential equations known as the Darboux system. Indeed, in the context of the geometric theory of integrable systems (see, e.g., \cite{RogersSchief2002} and references therein), the function $\psi$ constitutes an eigenfunction of the conjugate net equations (\ref{G2}) and the sets $\{\lambda^i_\alpha\}$ represent adjoint eigenfunctions. The functions $\psi_\alpha$ are Combescure transforms of the eigenfunction $\psi$ and, for reasons of symmetry, it is evident that each Combescure transform is a solution of another system of conjugate net equations with different coefficients.

\subsection{The generalised hodograph method}

In order to motivate the approach adopted in this paper, we here recall the generalised hodograph method developed by Tsarev in \cite{8} for a single system of hydrodynamic-type equations
\bela{G4a}
  u^i_{t} = \lambda^i(\bu)u^i_x,\qquad i = 1,\ldots, N + 1
\ela
with associated functions $A_{ik}$ defined by (\ref{G3}), that is,
\bela{G4b}
  \lambda^i_{u^k} = A_{ik}(\lambda^k - \lambda^i).
\ela
Thus, Tsarev's theorem states that if $\{\omega^i\}$ is another set of adjoint eigenfunctions obeying the above linear system then any local solution $\bu(x,t)$ of the nonlinear system
\bela{G4c}
  \omega^i(\bu) = \lambda^i(\bu)t + x
\ela
constitutes a solution of the hydrodynamic-type system (\ref{G4a}). Conversely, any solution of the hydrodynamic-type system may locally be represented in this manner. As indicated in the introduction, in the case $N=1$, (\ref{G4c}) may be regarded as a linear system for $x$ and $t$ rather than a nonlinear system for $u^1$ and $u^2$ and differentiation of $x(u^1,u^2)$ and $t(u^1,u^2)$ leads to the classical hodograph system \cite{CourantFriedrichs1948}
\bela{G4d}
  x_{u^1} + \lambda^2t_{u^1} = 0,\quad x_{u^2} + \lambda^1t_{u^2} = 0.
\ela
Here, the coefficients $\lambda^i$ are regarded as known functions of the independent variables $u^k$.
In the original context, this linear system is obtained from the nonlinear two-component system (\ref{G4a})$_{N=1}$ by merely interchanging dependent and independent variables, whereby the Jacobian determinant drops out.

\subsection{Generalised hodograph equations}

Even though the generalised hodograph method encapsulated in the algebraic system (\ref{G4c}) is applicable for all $N$, an associated system of hodograph-type equations is not available for \mbox{$N>1$} since the number of independent variables does not coincide with the number of dependent variables. However, since any flow which commutes with the hydrodynamic-type equations (\ref{G4a}) does not impose any constraint on the space of solutions, it is natural to supplement (\ref{G4a}) by $N-1$ commuting flows, leading to the larger system (\ref{G1}). Thus, if $\{\mu^i\}$ constitutes another set of adjoint eigenfunctions then we may locally define a coordinate transformation
\bela{G5}
  \bu = \bu(x,\bt)
\ela
via the system
\bela{G6}
  \mu^i(\bu) = \sum_{\alpha=1}^N\lambda^i_{\alpha}(\bu)t^{\alpha} + x
\ela
which coincides with the system (\ref{G4c}) in the case $N=1$.

It is now easy to see that Tsarev's generalised hodograph method is still valid in this more general setting so that, locally, the general solution of the hydrodynamic-type system (\ref{G1}) is encapsulated in the algebraic system (\ref{G6}) regarded as a definition of $\bu$. In fact, this observation may be interpreted as a corollary of Tsarev's theorem since if we select a ``time'' $t^{\alpha_0}$ and regard all other $t^{\alpha}$s as parameters then system (\ref{G6}) may be formulated as
\bela{G7}
  \omega^i(\bu) = \lambda^i_{\alpha_0}(\bu)t^{\alpha_0} + x,
\ela
where the quantities
\bela{G8}
  \omega^i = \mu^i - \sum_{\alpha\neq\alpha_0}\lambda^i_{\alpha}t^{\alpha}
\ela
represent linear superpositions of adjoint eigenfunctions, so that, according to the generalised hodograph method, (\ref{G1}) holds for $\alpha=\alpha_0$.

As in the classical case ($N=1$), the algebraic system (\ref{G6}) turns out to be equivalent to a system of first-order differential equations. Indeed, if we regard (\ref{G6}) as a definition of some functions $\mu^i$ then, on substitution into the adjoint eigenfunction equations  (\ref{G3}), it is readily verified that these functions constitute adjoint eigenfunctions if and only if the generalised hodograph equations
\bela{G10}
  x_{u^k}  + \sum_{\alpha=1}^N\lambda^i_{\alpha}t^\alpha_{u^k} = 0,\qquad i\neq k
\ela
are satisfied. By construction, this system of hodograph type is equivalent to the original hydrodynamic-type system (\ref{G1}).

\subsection{Iterated adjoint Darboux transformations}

It turns out that, just like the characteristic speeds $\lambda^i_\alpha$ and the quantities $\mu^i$, the remaining ingredients $x$ and $t^\alpha$ of the algebraic system (\ref{G6}) have distinct soliton-theoretic meaning. Thus, we first consider two sets $\{\mu^i\}$ and $\{\lambda^i\}$ of adjoint eigenfunctions obeying
\bela{G11}
  \mu^i_{u^k} = A_{ik}(\mu^k - \mu^i), \quad \lambda^i_{u^k} = A_{ik}(\lambda^k - \lambda^i)
\ela
for some solution $\{A_{ik}\}$ of the underlying Darboux system. This system is known to be invariant under adjoint Darboux transformations \cite{Eisenhart1962,KonopelchenkoSchief1993}. Specifically, for fixed $i$, the adjoint Darboux transformation $\Dbb^i$ generated by $\lambda^i$ transforms the adjoint eigenfunctions $\mu^l$ according to
\bela{G12}
 \bear{rl}
  \Dbb^i(\mu^i) = &\dis\mu^i - \frac{\lambda^i}{\lambda^i_{u^i}}\mu^i_{u^i}\AS
  \Dbb^i(\mu^k) = &\dis\frac{\lambda^i\mu^k - \lambda^k\mu^i}{\lambda^i - \lambda^k},\qquad k\neq i.
 \ear
\ela

By construction, the above Darboux transforms obey a linear system of the type (\ref{G11}) with coefficients depending on $A_{ik}$ and the adjoint eigenfunctions $\lambda^i$ only. The latter property guarantees that adjoint Darboux transformations may be iterated in the following purely algebraic manner. Given any $N$ sets of eigenfunctions $\{\lambda^i_{\alpha}\}$, we begin with the adjoint Darboux transformation $\Dbb^1_1$ generated by $\lambda^1_1$. The quantities $\Dbb^1_1(\mu^l)$ then constitute new adjoint eigenfunctions. In particular, if we focus on the new adjoint eigenfunctions $\Dbb^1_1(\lambda^l_2)$ then we may use the adjoint eigenfunction $\Dbb^1_1(\lambda^2_2)$ to define an adjoint Darboux transformation acting on the new adjoint eigenfunctions which we denote by $\Dbb^2_2$. This procedure may be repeated to construct $N$ adjoint Darboux transformations $\Dbb^1_1,\ldots,\Dbb^N_N$ generated by the adjoint eigenfunctions $\lambda^1_{1},\,\Dbb^1_1(\lambda^2_{2}),\,\Dbb^2_2(\Dbb^1_1(\lambda^3_3)),\ldots$. On use of Jacobi's identity for determinants \cite{Hirota2003}, it is then straightforward to verify by induction that the $N$th Darboux transform of the adjoint eigenfunction $\mu^{N+1}$ is given by
\bela{G13}
  (\Dbb^N_N\circ\cdots\circ\Dbb^1_1)(\mu^{N+1}) = \frac{\left|\bear{cccc}\lambda^1_{1}&\ldots&\lambda^1_{N}&\mu^1\\
                \vdots&&\vdots&\vdots\\ 
                \lambda^{N+1}_{1}&\ldots&\lambda^{N+1}_{N}&\mu^{N+1}
\ear\right|}{\left|\bear{cccc}\lambda^1_{1}&\ldots&\lambda^1_{N}&1\\
                \vdots&&\vdots&\vdots\\ 
                \lambda^{N+1}_{1}&\ldots&\lambda^{N+1}_{N}&1
\ear\right|}.
\ela
However, the right-hand side of the above expression is completely symmetric in both the upper and lower indices. Hence, the $N$th Darboux transform depends neither on the order of application of the adjoint Darboux transformations nor on the components of the sets of adjoint eigenfunctions $\{\lambda^i_\alpha\}$ which are chosen to generate the corresponding adjoint Darboux transformations. More precisely, for any permutations $(\alpha_1,\ldots,\alpha_N)$ and $(i_1,\ldots,i_{N+1})$ of $(1,\ldots,N)$ and \mbox{$(1,\ldots,N+1)$} respectively, the iterated Darboux transform
\bela{G14}
  (\Dbb^{i_N}_{\alpha_N}\circ\cdots\circ\Dbb^{i_1}_{\alpha_1})(\mu^{i_{N+1}}) = \mu_{(N)}
\ela
is the same. In fact, application of Cramer's rule shows that
\bela{G15}
 x = \mu_{(N)}
\ela
corresponds to the unique solution of the algebraic system (\ref{G6}) regarded as a linear system for $x$ and~$t^\alpha$. Thus, remarkably, by virtue of the commutativity of the flows (\ref{G1}), the ``spatial'' independent variable $x$ may be interpreted as the unique $N$-fold Darboux transform constructed from the characteristic speeds~$\lambda^i_\alpha$.

The interpretation of the ``times'' $t^\alpha$ is now based on the observation that the system (\ref{G6}) is implicitly symmetric in $x$ and $t^\alpha$. Indeed, for any fixed $\alpha$, the system (\ref{G6}) may be reformulated as
\bela{G17}
  \frac{\mu^i}{\lambda^i_{\alpha}} = \left(\sum_{\beta\neq\alpha}\frac{\lambda^i_{\beta}}{\lambda^i_{\alpha}}t^{\beta} + \frac{1}{\lambda^i_{\alpha}}x\right) + t^\alpha
\ela
so that the roles of $x$ and $t^\alpha$ have been interchanged. In fact, the {\em a priori} formal symmetry obtained in this manner may indeed be exploited by rewriting the linear system (\ref{G4}) as
\bela{G16}
  \psi_{u^i} = \frac{1}{\lambda^i_{\alpha}}\psi_{\alpha u^i},\quad \psi_{\beta u^i} = \frac{\lambda^i_{\beta}}{\lambda^i_{\alpha}}\psi_{\alpha u^i},\qquad \beta\neq\alpha.
\ela
The latter implies that the quantities $\tilde{\lambda}^i_\alpha=1/\lambda^i_\alpha$,  $\tilde{\lambda}^i_\beta=\lambda^i_\beta/\lambda^i_\alpha$ and, in fact, $\tilde{\mu}^i=\mu^i/\lambda^i_\alpha$ constitute adjoint eigenfunctions of the Darboux system associated with the eigenfunction $\psi_\alpha$. Hence, for reasons of symmetry, the time $t^\alpha$ obtained by means of Cramer's rule from (\ref{G17}) or, equivalently, the original system (\ref{G6}) coincides with the iterated Darboux transform
\bela{G18}
   t^{\alpha}  = \mu_{(N,\alpha)},\quad \mu_{(N,\alpha)} = (\tilde{\Dbb}^{i_N}_{\alpha_N}\circ\cdots\circ\tilde{\Dbb}^{i_1}_{\alpha_1})(\tilde{\mu}^{i_{N+1}}),
\ela
where the Darboux transformations $\tilde{\Dbb}^{i_k}_{\alpha_k}$ are now generated by the adjoint eigenfunctions $\tilde{\lambda}^{i_k}_{\alpha_k}$. It is also observed that the generalised hodograph system (\ref{G10}) may be solved for the derivatives of $t^\alpha$ to deduce that
\bela{G19}
  t^{\alpha}_{u^k} = \Lambda^\alpha_k(\lambda^i_{\beta})x_{u^k}
\ela
for some functions $\Lambda^\alpha_k$ and, hence, the $N+1$  variables $x$ and $t^\alpha$ may also be regarded as Combescure transforms of each other.

\section{Discrete generalised hodograph equations}

The formulation of the classical hodograph equations and their generalisation in the language of (adjoint) eigenfunctions may instantly be utilised to derive their canonical integrable discrete counterparts. Indeed, the standard integrable discretisation of the conjugate net equations (\ref{G2}) turns out to be the fundamental structure on which this discretisation technique is based. Thus, if 
\bela{G19a}
 \bear{c}
  \psi : \Z^{N+1}\rightarrow\R\as
  (n^1,\ldots,n^{N+1}) \mapsto\psi(n^1,\ldots,n^{N+1})
 \ear 
\ela
is an eigenfunction obeying the discrete conjugate net equations (see, e.g., \cite{11})
\bela{D17}
  \Delta_{ik}\psi = A_{ik}\Delta_i\psi + A_{ki}\Delta_k\psi,
\ela
where the forward difference operators $\Delta_i$ are defined by $\Delta_i\psi = \psi_{[i]} - \psi$ and
$\Delta_{ik} = \Delta_i\Delta_k = \Delta_k\Delta_i$, then a discrete Combescure transform $\psi_\star$ of $\psi$ defined~by
\bela{D18}
  \Delta_i\psi_\star = \lambda^i\Delta_i\psi
\ela
exists if the associated discrete adjoint eigenfunctions $\lambda^i$ constitute solutions of the linear system \cite{KonopelchenkoBogdanov95,DCN}
\bela{D18a}
  \Delta_k\lambda^i = A_{ik}(\lambda^k_{[i]} - \lambda^i_{[k]}).
\ela
Here, a subscript $_{[i]}$ denotes the relative unit increment $n^i\rightarrow n^i+1$ so that the mixed difference operator $\Delta_{ik}$ acts according to \mbox{$\Delta_{ik}\psi = \psi_{[ik]} - \psi_{[i]} - \psi_{[k]} + \psi$}. It is noted that (\ref{D17}) and (\ref{D18}) represent the discrete analogues of the linear equations (\ref{B2}) and (\ref{B4a}) defining the densities $\psi$ and fluxes $\psi_\star$ associated with the conservation laws for semi-Hamiltonian systems of hydrodynamic type. In connection with an appropriate Cauchy problem, it is convenient to reformulate the adjoint linear system (\ref{D18a}) as
\bela{D19}
  \Delta_k\lambda^i = \frac{A_{ik}}{A_{ik} + A_{ki}+1}(\lambda^k - \lambda^i).
\ela
As in the continuous case, the compatibility conditions for the (adjoint) eigenfunction equations (\ref{D17}) and (\ref{D18a}) (or (\ref{D19})) give rise to the same nonlinear system of discrete equations for the coefficients $A_{ik}$ which constitutes the standard integrable discretisation of the aforementioned Darboux system \cite{KonopelchenkoBogdanov95,DoliwaSantini97}. 

The discrete analogues of the classical adjoint Darboux transformations may be obtained by formally replacing derivatives by differences in the transformation laws (\ref{G12}). Indeed, the Darboux transforms of another set of adjoint eigenfunctions $\{\mu^l\}$ are given by
\bela{D20}
 \bear{rl}
  \Dbb^i(\mu^i) = &\dis\mu^i - \frac{\lambda^i}{\Delta_i\lambda^i}\Delta_i\mu^i = \frac{\lambda^i_{[i]}\mu^i - \lambda^i\mu^i_{[i]}}{\lambda^i_{[i]} - \lambda^i}\AS
  \Dbb^i(\mu^k) = &\dis\frac{\lambda^i\mu^k - \lambda^k\mu^i}{\lambda^i - \lambda^k},\qquad k\neq i
 \ear
\ela
for any fixed $i$ corresponding to the adjoint eigenfunction $\lambda^i$ which generates the adjoint Darboux transformation $\Dbb^i$. Since iteration of the discrete adjoint Darboux transformations only involves the algebraic transformation law (\ref{D20})$_2$ which coincides with the transformation law (\ref{G12})$_2$, the expressions (\ref{G13}), (\ref{G14}) and (\ref{G18})$_2$ for the iterated Darboux transforms are also valid in the discrete case. Moreover, the quantities $x$ and $t^\alpha$ defined~by
\bela{D21}
  x = \mu_{(N)},\quad t^{\alpha} = \mu_{(N,\alpha)}
\ela
still constitute the unique solution of the linear system (\ref{G6}), that is,
\bela{D22}
  \mu^i = \sum_{\alpha=1}^N\lambda^i_{\alpha}t^{\alpha} + x,
\ela
wherein $\lambda^i_\alpha$ and $\mu^i$ now refer to discrete adjoint eigenfunctions. In analogy with the continuous case, insertion into the adjoint eigenfunction equations (\ref{D19}) for $\{\mu^i\}$ leads to the linear system
\bela{D24}
  \Delta_kx  + \sum_{\alpha=1}^N\lambda^i_{\alpha [k]}\Delta_kt^\alpha = 0,\qquad i\neq k.
\ela
Conversely, any solution of this integrable discretisation of the generalised hodograph equations (\ref{G10}) provides via (\ref{D22}) a set of adjoint eigenfunctions $\{\mu^i\}$. Finally, the discrete generalised hodograph equations adopt the form
\bela{D25}
 \Delta_k t^{\alpha} = \Lambda^{\alpha}_k(\lambda^i_{\beta [k]})\Delta_kx,
\ela
which demonstrates that, as in the continuous case, the $N+1$ variables $x$ and~$t^\alpha$ may be interpreted as discrete Combescure transforms of each other.

As pointed out in the previous section, there exists complete equivalence between the hydrodynamic-type system (\ref{G1}) and the generalised hodograph equations (\ref{G10}). In fact, this is verified directly by employing a formulation in terms of differential forms (cf.\ \cite{Ferapontov02}). Indeed, it is seen that the system
\bela{D26}
  du^i\wedge dx + \sum_{\alpha=1}^N\lambda^i_\alpha\, du^i\wedge dt^\alpha = 0
\ela
reduces to the hydrodynamic-type system (\ref{G1}) if $x$ and $t^\alpha$ are chosen as the independent variables. Alternatively, one may select the $u^i$s as the independent variables so that the generalised hodograph equations (\ref{G10}) result. Accordingly, the algebraic system (\ref{G6}) encodes the \mbox{$N+1$}-dimensional integral manifolds $\mathcal{M}$ of the differential system (\ref{D26}). Thus, if we interpret a solution $(x,\bt)(\bu)$ of the generalised hodograph equations (\ref{G10}) as an $N+1$-dimensional submanifold $\mathcal{M}$ of the space of (in)dependent variables $\R^{2N+2}$ then this submanifold $\mathcal{M}$ admits the parametrisation
\bela{D13}
  \bu \mapsto (x(\bu),\bt(\bu),\bu).
\ela
However, locally, we may also utilise the parametrisation
\bela{D13a}
  (x,\bt) \mapsto (x,\bt,\bu(x,\bt)),
\ela
where $\bu(x,\bt)$ represents a corresponding solution of the hydrodynamic-type system~(\ref{G1}). In the discrete case, the algebraic system (\ref{D22}) encapsulates ``discrete integral manifolds'' $\mathcal{M}^\Delta$ in the following sense. Any solution 
\bela{D15}
  (x,\bt) : \Z^{N+1}\rightarrow \R^{N+1},\quad \n\mapsto (x,\bt)(\n)
\ela
of the discrete generalised hodograph equations (\ref{D24}) may be used to parametrise a ``discrete submanifold'' $\mathcal{M}^\Delta$ of $\R^{N+1}\times\delta^1\Z\times\cdots\times\delta^{N+1}\Z$ according to
\bela{D16}
  \n \mapsto (x(\n),\bt(\n),\bu^\Delta),
\ela
where $\bu^\Delta = (\delta^1n^1,\ldots,\delta^{N+1}n^{N+1})$ for prescribed lattice parameters $\delta^i$. 
Hence, the variable $\bu^\Delta$ may be regarded as a discretisation of either the independent variables of the generalised hodograph equations (\ref{G10}) or the dependent variables of the system of hydrodynamic type (\ref{G1}). The latter corresponds to an ``implicit discretisation'' of the hydro\-dy\-namic-type system with variable spacing between the lattice points on the $N+1$-dimensional submanifold $\R^{N+1}$ of the independent variables $x$ and $\bt$.

\section{Discrete Euler-Poisson-Darboux systems}

It is well known \cite{13,14,15,16} that the characteristic speeds $\lambda^i$ associated with the multi-phase averaged Korteweg-de Vries (KdV) equations are related to linear hyperbolic equations of Euler-Poisson-Darboux type. In fact, recently, it has been demonstrated \cite{18} that these characteristics speeds may be generated by means of iterated Darboux transformations applied to separable solutions of (extended) Euler-Poisson-Darboux-type systems. It turns out that one may construct canonical discretisations of the multi-phase characteristic speeds if one carefully defines analogues of the hyperelliptic integrals associated with the underlying Riemann surfaces of genus $g\geq1$. In this section, we demonstrate how one may derive particular classes of discrete characteristic speeds from the discrete Euler-Poisson-Darboux-type system
\bela{E1}
  [\delta^i(n^i+\nu^i) - \delta^k(n^k+\nu^k)]\Delta_{ik}\phi = \delta^k\epsilon^k\Delta_i\phi - \delta^i\epsilon^i\Delta_k\phi,
\ela
where $i\neq k\in\{1,\ldots,2g+1\}$, which  include those of ``averaged KdV'' type. Here, the constants $\delta^i$ are lattice parameters and the constants $\epsilon^i$ determine the nature of the contour integrals to be defined in \S 5. For $\epsilon^i = \frac{1}{2}$, this leads to analogues of the above-mentioned hyperelliptic integrals. The parameters $\nu^i$ reflect the fact that it is crucial to maintain the freedom of placing the discretisation points not necessarily on the vertices of the $\Z^{2g+1}$ lattice but, possibly, on the edges, faces etc. Thus, we regard the hyperbolic system (\ref{E1}) as a discretisation of the classical Euler-Poisson-Darboux system
\bela{E1a}
  (x^i - x^k)\phi_{x^ix^k} = \epsilon^k\phi_{x^i} - \epsilon^i\phi_{x^k}
\ela
obtained in the limit $x^i = \delta^i (n^i + \nu^i)$, $\delta^i\rightarrow0$. In the following, the key idea is to introduce an auxiliary continuous variable $y$ and supplement the discrete Euler-Poisson-Darboux system by the differential-difference equations
\bela{E1aa}
  [y - \delta^i (n^i+\nu^i)]\Delta_i\phi_y = \delta^i\epsilon^i\phi_y + (g-1)\Delta_i\phi.
\ela
The function $\phi = \phi(\n,y)$ is well-defined since the semi-discrete Euler-Poisson-Darboux system (\ref{E1}), (\ref{E1aa}) remains compatible. 

\subsection{Separable solutions}

As in the continuous case \cite{18}, we now focus on separable solutions of the semi-discrete Euler-Poisson-Darboux system (\ref{E1}), (\ref{E1aa}). Thus, it is readily verified that the ansatz
\bela{E1b}
  \phi_{\rm sep} = \rho(\zeta)\prod_{i=1}^{2g+1}\rho^{i}(\zeta),
\ela
where we have suppressed the dependence of $\rho$ and $\rho^i$ on $y$ and $n^i$ respectively, leads to the first-order differential/difference equations
\bela{E1c}
  \Delta_i\rho^{i} = \frac{\delta^i\epsilon^i}{\zeta - \delta^i(n^i+\nu^i)}\rho^{i},\quad \rho_y = \frac{(1-g)}{\zeta - y}\rho
\ela
with $\zeta$ being a (complex) constant of separation. The latter may be solved to obtain
\bela{E1d}
  \rho = (\zeta - y)^{g-1}
\ela
without loss of generality and, in the continuum limit $\delta^i\rightarrow0$, the difference equations (\ref{E1c})$_1$ reduce to 
\bela{E1da}
  \rho_{x^i}^i = \frac{\epsilon^i}{\zeta - x^i}\rho^i.
\ela
Hence, up to a multiplicative constant, $\rho^{i}$ represents a canonical discretisation of $(\zeta - x^i)^{-\epsilon^i}$ so that
\bela{E1i}
  \phi_{\rm sep} = \rho(\zeta)\prod_{i=1}^{2g+1}\rho^{i}(\zeta)\quad\rightarrow\quad(\zeta-y)^{g-1}\prod_{i=1}^{2g+1}(\zeta - x^i)^{-\epsilon^i}
\ela
in the continuum limit.

\subsection{Superposition and iterated Darboux transformations}

The separable solutions derived in the preceding may be superimposed to obtain large classes of solutions of the semi-discrete Euler-Poisson-Darboux system (\ref{E1}), (\ref{E1aa}). Here, we consider the contour integrals 
\bela{E3}
  \phi^\kappa = \oint_{b_\kappa}(\zeta-y)^{g-1}\prod_{i=1}^{2g+1}\rho^i(\zeta)\,d\zeta,\qquad \kappa = 1,\ldots, g,
\ela
where the contours $b_\kappa$ on the complex $\zeta$-plane are assumed to be independent of $y$ and ``locally'' independent of $\n$, that is, we demand that
\bela{E3a}
   \Delta_i\oint_{b_\kappa}f(\n,y;\zeta)\,d\zeta = \oint_{b_\kappa}\Delta_if(\n,y;\zeta)\,d\zeta
\ela
for any relevant functions $f$. Accordingly, the semi-discrete Euler-Poisson-Darboux system (\ref{E1}), (\ref{E1aa}) admits vector-valued solutions of the form
\bela{E2}
  \phib = \left(\bear{c} \phi^1\\ \vdots\\\phi^g\ear\right),\quad
  \hat{\phib} = \left(\bear{c} \phi^2\\ \vdots\\\phi^{g}\ear\right).
\ela
The components of these solutions may be used to generate iteratively solutions of semi-discrete conjugate net equations with increasingly complex coefficients. Thus, the $(g-1)$-fold Darboux transform \cite{IDT} of any solution $\phi$ of the semi-discrete Euler-Poisson-Darboux system (\ref{E1}), (\ref{E1aa}) with respect to the independent variable $y$ is given by the compact expression
\bela{E4}
  \phi_{g-1} = \frac{\left|\bear{cccc}\phi & \phi_y & \cdots & \phi_{(g-1)y}\\
                                                      \hat{\phib} & \hat{\phib}_y & \cdots & \hat{\phib}_{(g-1)y}\ear\right|}{\left|\bear{ccc}\hat{\phib}_y & \cdots & \hat{\phib}_{(g-1)y}\ear\right|}.
\ela
In the context of classical differential geometry, this Darboux transform is known as the \mbox{$(g-1)$}-fold Levy transform \cite{Eisenhart1962} with respect to $y$. The Levy transforms of the particular solutions
\bela{E5}
  \phi^0 = \sum_{i=1}^{2g+1}\epsilon^i\delta^in^i - (g-1)y, \quad \phi^1
\ela
of the semi-discrete Euler-Poisson-Darboux system therefore read
\bela{E6}
 \bear{rl}
  \phi^0_{g-1}= &\dis\frac{\sum_{i=1}^{2g+1}\epsilon^i\delta^i n^i\,|\hat{\phib}_y\,\,\cdots\,\,\hat{\phib}_{(g-1)y}| + (g-1)|\hat{\phib} - y\hat{\phib}_y\,\,\hat{\phib}_{yy}\,\,\cdots\,\,\hat{\phib}_{(g-1)y}|}{|\hat{\phib}_y\,\,\cdots\,\,\hat{\phib}_{(g-1)y}|}\AS
  \phi^1_{g-1}=&\dis\frac{|\phib\,\,\cdots\,\,\phib_{(g-1)y}|}{|\hat{\phib}_y\,\,\cdots\,\,\hat{\phib}_{(g-1)y}|}
 \ear
\ela
so that it is readily verified that
\bela{E7}
 \bear{rl}
  \phi^0_{g-1} = &\dis\frac{\sum_{i=1}^{2g+1}\epsilon^i\delta^in^i\,|\hat{\Ib}_{g-2}\,\,\cdots\,\,\hat{\Ib}_0| - |\hat{\Ib}_{g-1}\,\,\hat{\Ib}_{g-3}\,\,\cdots\,\,\hat{\Ib}_0|}{|\hat{\Ib}_{g-2}\,\,\cdots\,\,\hat{\Ib}_0|}\AS
   \phi^1_{g-1}=&\dis\frac{|\Ib_{g-1}\,\,\cdots\,\,\Ib_0|}{|\hat{\Ib}_{g-2}\,\,\cdots\,\,\hat{\Ib}_0|}
 \ear
\ela
with the contour integrals
\bela{E8}
 I_k^\kappa = \oint_{b_\kappa}\zeta^k\prod_{i=1}^{2g+1}\rho^i(\zeta)\,d\zeta.
\ela
It is observed that, remarkably, the Levy transforms $\phi^0_{g-1}$ and $\phi^1_{g-1}$ are independent of $y$ and that, by definition, $\phi^0_0=\phi^0$ and $\phi^1_0=\phi^1$ in the case $g=1$.

The action of another Levy transformation with respect to the variable $n^i$ now produces the $g$-fold Levy transform
\bela{E9}
  \lambda^i = \phi^i_g[\phi^0] = \phi^0_{g-1} - \frac{\phi^1_{g-1}}{\Delta_i\phi^1_{g-1}}\Delta_i\phi^0_{g-1}
\ela
of $\phi^0$. Here, the symbol $\lambda^i$ has been chosen to indicate that the set $\{\lambda^i\}$ will indeed be shown to
constitute a set of adjoint eigenfunctions. Since the Levy transform $\lambda^i$ may be formulated as
\bela{E10}
  \lambda^i = \frac{\Delta_i(\phi^0_{g-1}/\phi^1_{g-1})}{\Delta_i(1/\phi^1_{g-1})},
\ela
we may set
\bela{E11}
 \bear{c}\dis
  H_1 = \frac{|\hat{\Ib}_0\,\,\cdots\,\,\hat{\Ib}_{g-2}|}{|\Ib_0\,\,\cdots\,\,\Ib_{g-1}|},\quad
  H_2 = -\frac{|\hat{\Ib}_0\,\,\cdots\,\,\hat{\Ib}_{g-3}\,\,\hat{\Ib}_{g-1}|}{|\Ib_0\,\,\cdots\,\,\Ib_{g-1}|}\AS
 \dis \bar{\Gamma}_0 = 1,\quad \bar{\Gamma}_1 = \sum_{i=1}^{2g+1}\epsilon^i\delta^in^i
 \ear
\ela
to obtain the final expression
\bela{E12}
  \lambda^i = \frac{\Delta_i(H_1\bar{\Gamma}_1 + H_2\bar{\Gamma}_0)}{\Delta_iH_1}.
\ela
This constitutes a natural discretisation of a particular case of the characteristic speeds obtained  in an entirely different manner by Tian in the continuous context (cf.\ \cite[p.\ 218]{16} for $\alpha_1=1$, $\alpha_k=0$ otherwise and $H_{1/2}\sim K^{(1)}_{1/2}$ in Tian's notation). Once again, in the case $g=1$, the interpretation $H_1=1/I^1_0$ and $H_2=0$ is to be adopted.

\subsection{Discrete characteristic speeds}

The connection with discrete characteristic speeds and the associated discrete generalised hodograph equations (\ref{D24}) is now made as follows. By construction, $\phi^0_{g-1}$ and $\phi^1_{g-1}$ are solutions of the same system of discrete conjugate net equations
\bela{E13}
  \Delta_{ik}\phi_{g-1} = B_{ik}\Delta_i\phi_{g-1} + B_{ki}\Delta_k\phi_{g-1}.
\ela
On the other hand, the classical Levy transforms of an eigenfunction corresponding to different ``directions'' $x^i$ may also be regarded as adjoint eigenfunctions of another system of conjugate net equations \cite{KonopelchenkoSchief1993}. The analogous statement is true in the discrete case and, accordingly, the quantities $\lambda^i$ constitute adjoint eigenfunctions associated with the discrete conjugate net equations
\bela{E14}
  \Delta_{ik}\psi = A_{ik}\Delta_i\psi + A_{ki}\Delta_k\psi,
\ela
wherein the coefficients $A_{ik}$ are related to the coefficients $B_{ik}$ by
\bela{E16}
  C_{ik} = \frac{A_{ik}}{A_{ik} + A_{ki} + 1} = B_{ki}\frac{\phi^1_{g-1[k]}\Delta_k\phi^1_{g-1}}{\phi^1_{g-1}\Delta_i\phi^1_{g-1[k]}} - \frac{\Delta_k\phi^1_{g-1}}{\phi^1_{g-1}}.
\ela
The above observation allows us to identify a particular set of discrete characteristic speeds which may be used in the discrete generalised hodograph equations (\ref{D24}). However, the definition of the latter requires $2g+1$ sets of adjoint eigenfunctions $\{\lambda^i_\alpha\}$, each of which represents a solution of the adjoint eigenfunction equations
\bela{E15}
  \Delta_k\lambda^i = C_{ik}(\lambda^k - \lambda^i)
\ela
satisfied by the Levy transforms $\lambda^i$. 

In order to construct canonical sets of adjoint eigenfunctions satisfying (\ref{E15}), it is required to introduce an explicit parametrisation of the functions $\rho^i$ in the base separable solution (\ref{E1b}) of the associated semi-discrete Euler-Poisson-Darboux system. Thus, in terms of Gamma functions \cite{AbramowitzStegun1964}, the general solution of the difference equation (\ref{E1c})$_1$ formulated as 
\bela{E15a}
  \rho^i_{[i]} = \frac{\zeta - \delta^i (n^i+\nu^i - \epsilon^i)}{\zeta - \delta^i(n^i+\nu^i)}\rho^i
\ela
is given by
\bela{E15b}
  \rho^i = {(\delta^i)}^{-\epsilon^i}\frac{\Gamma(\xi^i - n^i - \nu^i + 1)}{\Gamma(\xi^i - n^i - \nu^i + \epsilon^i + 1)},\qquad \xi^i = \frac{\zeta}{\delta^i}
\ela
up to a constant of ``integration'' which may depend on $\zeta$. In fact, the multiplicative factor has been chosen in such a manner that $\rho^i\rightarrow (\zeta - x^i)^{-\epsilon^i}$ in the continuum limit $\delta^i\rightarrow0$. This is a consequence of the well-known asymptotic behaviour
\bela{E15c}
  \lim_{|z|\rightarrow\infty} z^{b-a}\frac{\Gamma(z + a)}{\Gamma(z+b)} = 1,\quad |\operatorname{arg}(z)|<\pi
\ela
of ratios of Gamma functions. In fact, the first two terms of the associated classical asymptotic expansion \cite{TricomiErdelyi1951} read
\bela{P5a}
  \frac{\Gamma(z+a)}{\Gamma(z+b)} = z^{a-b}\left[1 + \frac{(a-b)(a+b -1)}{2z} + O(|z|^{-2})\right].
\ela
It is noted that (\ref{E15b}) regarded as a discretisation of a ``power function'' essentially coincides with that considered in \cite{GelfandGraevRetakh92}.  

It has been pointed out that the Levy transforms $\lambda^i=\phi^i_g[\phi^0]$ are independent of the auxiliary variable $y$. This is due to the fact that the seed solution $\phi^0$ of the semi-discrete Euler-Poisson-Darboux system is a polynomial in $y$ of degree at most $g-1$. A canonical way of generating an infinite number of seed solutions which admit this property is to expand the separable solution
\bela{E16b}
 \phi  = \zeta^\sigma\phi_{\rm sep} = \zeta^\sigma(\zeta - y)^{g-1}\prod_{i=1}^{2g+1}\rho^i(\zeta),\qquad
\sigma= \sum_{i=1}^{2g+1}\epsilon^i - (g-1)
\ela
about $\zeta=\infty$ to obtain
\bela{E16c}
  \phi = (1 - y \zeta^{-1})^{g-1}\sum_{m=0}^{\infty}\Gamma_m(\n)\zeta^{-m}.
\ela
The existence of this formal power series in $\zeta^{-1}$ is readily established by applying the asymptotic expansion (\ref{P5a}) to the function $\rho^i$ as given by (\ref{E15b}) and reformulating it as an asymptotic series in $\zeta^{-1}$, namely 
\bela{E16cc}
  \rho^i(\zeta) = \zeta^{-\epsilon^i}\left[1 + \epsilon^i\delta^i\left(n^i + \nu^i - \frac{\epsilon^i+1}{2}\right)\zeta^{-1} + O(|\zeta|^{-2})\right].
\ela
Thus, for instance, the first two coefficients $\Gamma_0$ and $\Gamma_1$ are seen to be
\bela{E16ccc}
  \Gamma_0 = 1,\quad \Gamma_1 = \sum_{i=1}^{2g+1}\epsilon^i\delta^i\left(n^i  + \nu^i - \frac{\epsilon^i+1}{2}\right).
\ela
It is evident that the expansion (\ref{E16c}) is of the form
\bela{E24}
  \phi = \sum_{\alpha=0}^{\infty}\Xi_\alpha(\n,y)\zeta^{-\alpha},
\ela
where the coefficients $\Xi_\alpha(\n,y)$ are polynomials in $y$ of degree $\alpha$ if $\alpha\leq g-1$ and of degree $g-1$ if $\alpha> g-1$. In fact, 
\bela{E25}
  \Xi_\alpha(\n,y) = \sum_{k=0}^{g-1}(-y)^k\binom{g-1}{k}\Gamma_{\alpha,k}(\n),
\ela
where
\bela{E26}
  \Gamma_{\alpha,k} = \Gamma_{\alpha-k}\quad\mbox{ if }\quad 0\leq k \leq \operatorname{min}(\alpha,g-1)
\ela
and $\Gamma_{\alpha,k} = 0$ otherwise. By construction, each coefficient $\Xi_\alpha$ constitutes a solution of the semi-discrete Euler-Poisson-Darboux system (\ref{E1}), (\ref{E1aa}). For instance, $\Xi_0 = \Gamma_0=1$ represents the trivial constant solution, while
\bela{E27}
  \Xi_1 = \Gamma_1 - (g-1)y\Gamma_0  = \phi^0 + c^0
\ela
turns out to be a linear superposition of the trivial solution and the seed solution $\phi^0$ which has been used to construct the discrete characteristic speeds $\lambda^i$ given by~(\ref{E12}). The constant $c^0$ may be read off (\ref{E16ccc}).

The general  expression (\ref{E4}) for the iterated Darboux transform $\phi_{g-1}$ may be used to generate the $(g-1)$-fold Levy transform $\phi_{g-1}[\Xi_\alpha]$ of any seed solution $\Xi_\alpha$. Since the degree of $\Xi_\alpha$ in $y$ is less than $g$, the Levy transform $\phi_{g-1}[\Xi_\alpha]$ is independent of $y$. Hence, the procedure outlined in \S 4(b) may be simplified by evaluating the analogue of (\ref{E6}) at $y=0$. As a result, one is immediately led to the compact expression
\bela{E28}
  \phi_{g-1}[\Xi_\alpha] =  \frac{\left|\bear{cccc}\Gamma_{\alpha,0} & \Gamma_{\alpha,1} & \cdots & \Gamma_{\alpha,g-1}\\ \hat{\Ib}_{g-1} & \hat{\Ib}_{g-2} & \cdots & \hat{\Ib}_0\ear\right|}{|\hat{\Ib}_{g-2}\,\,\cdots\,\,\hat{\Ib}_0|}.
\ela
In particular, by virtue of (\ref{E27}), it may be concluded that
\bela{E29}
  \phi_{g-1}[\Xi_1] = \phi_{g-1}^0 + c^0
\ela
so that, essentially, the $(g-1)$-fold Levy transform associated with the discrete characteristic speeds $\lambda^i$ is retrieved. We may now employ the eigenfunctions $\phi_{g-1}[\Xi_\alpha]$ and $\phi_{g-1}^1$ to generate additional discrete characteristic speeds in the manner described in \S 4(b) by replacing $\phi_{g-1}^0$ by $\phi_{g-1}[\Xi_\alpha]$ in (\ref{E9}) and (\ref{E10}). In terms of the coefficients $\Gamma_{\alpha,k}$ and the ratios of determinants
\bela{E31}
H_k = (-1)^{k+1}\frac{|\hat{\Ib}_0\,\,\cdots\,\,\hat{\Ib}_{g-k-1}\,\,\hat{\Ib}_{g-k+1}\,\,\cdots\,\,\hat{\Ib}_{g-1}|}{|\Ib_0\,\,\cdots\,\,\Ib_{g-1}|},
\ela
these turn out to be
\bela{E30}
  \lambda^i_\alpha = \phi^i_g[\Xi_\alpha] =\frac{\Delta_i(H_1\Gamma_{\alpha,0} + \cdots + H_g\Gamma_{\alpha,g-1})}{\Delta_iH_1}
\ela
and encode the discrete characteristic speeds $\lambda^i$ via
\bela{E30a}
  \lambda^i_1 = \lambda^i + c^0.
\ela
Accordingly, any choice of $2g+1$ sets of adjoint eigenfunctions $\{\lambda^i_\alpha\}$ such as \mbox{$\alpha = 1,\ldots, 2g+1$} gives rise to a discrete system of generalised hodograph equations (\ref{D24}) with associated ``implicitly defined'' discrete commuting flows of hydrodynamic type. Once again, it is observed that (\ref{E30}) represents a natural discretisation of the compact formulation of the corresponding characteristic speeds recorded in \cite{16}.

\section{``Discrete'' hyperelliptic integrals and characteristic speeds of Whitham type}

It has been demonstrated that the characteristic speeds $\lambda^i_\alpha$ are independent of the auxiliary variable $y$ and, accordingly, the contour integrals
\bela{E30b}
 I_k^\kappa = \oint_{b_\kappa}\zeta^k\prod_{i=1}^{2g+1}\rho^i(\zeta)\,d\zeta
\ela
constitute the main ingredients in their construction. Here, we are concerned with contours $b_\kappa$ which mimic canonical cycles associated with classical hyperelliptic integrals \cite{hyper}. To this end, we make the choice
\bela{E30c}
  \epsilon^i = \frac{1}{2},\quad \delta^i = \delta
\ela
so that the underlying discrete Euler-Poisson-Darboux system (\ref{E1}) reduces to 
\bela{H1}
  2(n^i + \nu^i - n^k - \nu^k)\Delta_{ik}\phi = \Delta_i\phi - \Delta_k\phi
\ela
and the continuum limit is represented by $\delta\rightarrow 0$ with $\delta n^i$ in $x^i = \delta(n^i + \nu^i)$ held constant as before. Up to the factor $\zeta^k$, the integrand of the contour integrals (\ref{E30b}) may be formulated as
\bela{H2}
  \varphi = \prod_{i=1}^{2g+1}p(\xi,n^i,\nu^i),\quad\xi = \frac{\zeta}{\delta},
\ela
where the function $p$ representing all functions $\rho^i$ is defined by
\bela{H3}
  p(\xi,n,\nu) = \frac{\Gamma(\xi - n - \nu + 1)}{\sqrt{\delta}\,\Gamma(\xi - n - \nu + \frac{3}{2})}
\ela
in agreement with the choice (\ref{E15b}).

\subsection{``Discrete'' cycles and hyperelliptic integrals}

\begin{figure}
\centerline{\includegraphics[scale=0.4]{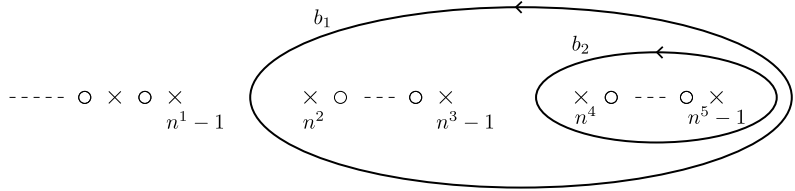}}
\caption{The distribution of zeros (circles) and poles (crosses) of the function $\varphi$ on the $\xi$-plane and associated $b$-cycles for $g=2$.}
\label{cycles}
\end{figure}

We now introduce the ordering
\bela{H4}
  n^1 < n^2 < \cdots < n^{2g} < n^{2g+1}
\ela
and choose 
\bela{H5}
  \nu^{2k} = \frac{1}{2},\quad \nu^{2k+1} = 0
\ela
corresponding to the discretisation points $x^{2k} = \delta(n^{2k} +\frac{1}{2})$ and $x^{2k+1} = \delta n^{2k+1}$. Hence, the separable solution (\ref{H2}) of the discrete Euler-Poisson-Darboux system (\ref{H1}) becomes
\bela{H6}
 \varphi = \frac{1}{\delta^{g+\frac{1}{2}}}
    \prod_{k=0}^{g}\frac{\Gamma(\xi - n^{2k+1} + 1)}{\Gamma(\xi - n^{2k+1} + \frac{3}{2})}
    \prod_{k=1}^{g}\frac{\Gamma(\xi - n^{2k} + \frac{1}{2})}{\Gamma(\xi - n^{2k} + 1)}.
\ela
For instance, in the case $g=1$, we obtain
\bela{H7}
  \varphi = \frac{1}{\delta^{\frac{3}{2}}}
                \frac{\Gamma(\xi - n^1 + 1)}{\Gamma(\xi - n^1 + \frac{3}{2})}
                \frac{\Gamma(\xi - n^2 + \frac{1}{2})}{\Gamma(\xi - n^2 + 1)}
                \frac{\Gamma(\xi - n^3 + 1)}{\Gamma(\xi - n^3 + \frac{3}{2})}.
\ela
Since the Gamma function is non-zero but has simple poles at non-positive integers, the distribution of zeros and poles of the function $p(\xi,n,0)$ is given by
\bela{H8}
 \bear{rll}
  p(\xi,n,0) & = 0,&\quad \xi = \ldots,n-\frac{5}{2},n-\frac{3}{2}\as
  p(\xi,n,0) & = \pm\infty,&\quad \xi = \ldots,n-2,n-1,
 \ear
\ela
whereas
\bela{H9}
 \bear{rll}
  p(\xi,n,\frac{1}{2}) & = 0,&\quad \xi = \ldots,n-2,n-1\as
  p(\xi,n,\frac{1}{2}) & = \pm\infty,&\quad \xi = \ldots,n-\frac{3}{2},n-\frac{1}{2}.
 \ear
\ela
Accordingly, the zeros and poles of the functions $p$ which make up $\varphi$ partially cancel each other in such a manner that, as a function of $\xi$, $\varphi$ has no zeros or poles in the region
\bela{H10}
  \bigcup_{k=1}^{g+1}(n^{2k-1}-1,n^{2k}),\qquad n^{2g+2} = \infty.
\ela
It is therefore natural to define the $g$ contours $b_{\kappa}$ (on the $\xi$-plane) as closed paths of counterclockwise orientation which pass through the pairs of intervals $(n^{2\kappa-1}-1,n^{2\kappa})$ and $(n^{2g+1}-1,\infty)$ for $\kappa=1,\ldots,g$ as indicated in Figure \ref{cycles}. 
The contour integrals 
\bela{H10a}
  \oint_{b_\kappa}\zeta^k\varphi(\xi)\,d\zeta,\qquad \kappa=1,\ldots,g,\quad k=0,\ldots,g-1
\ela
constitute ``discrete'' analogues of the hyperelliptic integrals
\bela{H10b}
  \oint_{b_\kappa}\frac{\zeta^k}{\sqrt{\prod_{i=1}^{2g+1}(\zeta - x^i)}}d\zeta
\ela
with the contours $b_\kappa$ essentially becoming the $b$-cycles employed in \cite{16,17} in the limit $\delta\rightarrow0$. The intervals $(n^{2\kappa-1}-1,n^{2\kappa})$ and $(n^{2g+1}-1,\infty)$, $\kappa=1,\ldots,g$ correspond to the cuts $(x^{2\kappa-1},x^{2\kappa})$ and $(x^{2g+1},\infty)$ along which the upper and lower sheets of the underlying Riemann surface of genus $g$ are joined. In fact, the $\zeta$-plane represents the union of the half of the upper sheet and the half of the lower sheet which contain the $b$-cycles. This union is discontinuous between the cuts and, in the discrete case, this is reflected by the presence of poles and zeros between the intervals $(n^{2\kappa-1}-1,n^{2\kappa})$ and $(n^{2g+1}-1,\infty)$. As in the continuous case, the ``discrete'' $b$-cycles are ``locally'' independent of $\n$ in the sense of (\ref{E3a}) so that the ``discrete'' hyperelliptic integrals (\ref{H10a}) regarded as functions of $\n$ are indeed solutions of the discrete Euler-Poisson-Darboux system (\ref{E1}). 

The discrete hyperelliptic integrals (\ref{H10a}) may be evaluated explicitly in terms of the residues of the meromorphic integrand $\varphi$ since one only requires the known relationships
\bela{H10c}
  \Gamma({\textstyle\frac{1}{2}} \pm l) = \left[\frac{(2l)!}{(\pm 4)^l l!}\right]^{\pm1}\sqrt{\pi},\quad 
  \operatorname{res}(\Gamma(z),z=-l) = \frac{(-1)^l}{l!},\quad l\in\mathbbm{N}.
 \ela
Specifically, in the case $g=1$, the contour integral
\bela{H11}
  \phi^\Delta = \frac{1}{2\pi i}\oint_{b_1}\varphi(\xi)\,d\zeta = \frac{\delta}{2\pi i}\oint_{b_1}\varphi(\xi)\,d\xi=\delta \sum_{k=n^2}^{n^3-1}\operatorname{res}(\varphi(\xi),\xi=k)
\ela
is given by
\bela{H13}
  \phi^\Delta = \frac{1}{\sqrt{\delta}}\sum_{k=n^2}^{n^3-1} 
                \frac{\Gamma(k - n^1 + 1)}{\Gamma(k - n^1 + \frac{3}{2})}
                \frac{\Gamma(k - n^2 + \frac{1}{2})}{\Gamma(k - n^2 + 1)}
                \frac{\operatorname{res}(\Gamma(\eta),\eta = k - n^3 + 1)}{\Gamma(k - n^3 + \frac{3}{2})}.
\ela
This is to be compared with the corresponding elliptic integral (\ref{H10b}) (divided by $2\pi i$) evaluated at the points
\bela{H13a}
   x^1 = \delta n^1,\quad x^2 = \delta(n^2 + \textstyle\frac{1}{2}),\quad x^3 = \delta n^3.
\ela
In terms of the complete elliptic integral $\sfK$ of the first kind \cite{AbramowitzStegun1964}, this elliptic integral may be expressed as
\bela{H12}
  \phi^{\rm O} = \frac{2}{\pi\sqrt{x^3-x^1}}\sfK\left(\sqrt{\frac{x^3-x^2}{x^3-x^1}}\right).
\ela
By construction, the latter constitutes an eigenfunction of the continuous Euler-Poisson-Darboux system (\ref{E1a}) and may be used to generate three adjoint eigenfunctions $\lambda^1,\lambda^2,\lambda^3$ by means of the continuous analogue of the Levy transformation (\ref{E10}) for $g=1$. These turn out to be the characteristic speeds in the one-phase averaged KdV equations derived by Whitham \cite{2,8}. Thus, the discrete elliptic integral $\phi^{\Delta}$ gives rise to discrete characteristic speeds of Whitham type. 

\begin{figure}
\centerline{\includegraphics[width=0.5\textwidth]{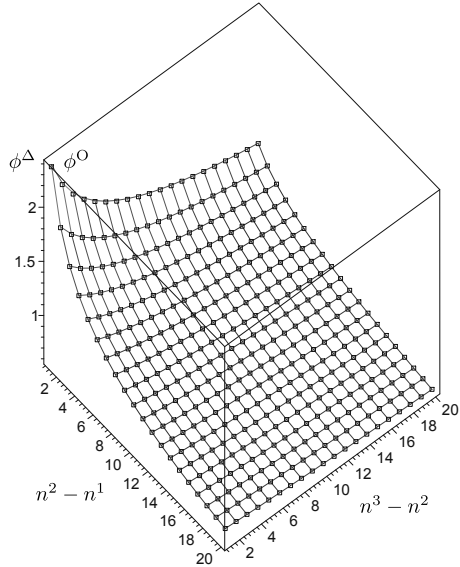}}
\caption{The (discrete) elliptic integrals $\phi^\Delta$ (points) and $\phi^{\rm O}$ (mesh) plotted as functions of the differences $n^2-n^1$ and $n^3-n^2$.}
\label{riemann}
\end{figure}

It is observed that the elliptic integral $\phi^{\rm O}$ only depends on the differences of the $x^i$. In fact, the same applies, {\em mutatis mutandis}, to the discrete elliptic integral (\ref{H11}) since the function $p(\xi,n,\nu)$ is invariant under a shift of $\xi$ and $n$ by the same amount. Accordingly, it is natural to regard the (discrete) elliptic integrals $\phi^\Delta$ and $\phi^{\rm O}$ as functions of the differences $n^2-n^1$ and \mbox{$n^3-n^2$}. Their graphs are displayed in Figure \ref{riemann} and it its seen that there exists virtually no difference between the discrete and continuous elliptic integrals represented by points and a mesh respectively.
It is noted that this statement is independent of the lattice parameter $\delta$ in the sense that, as a function of $\n$, the ratio $\phi^\Delta/\phi^{\rm O}$ does not depend on $\delta$. Thus, remarkably, there exists a unique relationship between the discrete and continuous elliptic integrals. We conclude with the remark that the summation involved in the determination of the discrete hyperelliptic integrals may be reformulated so that it becomes transparent that the discrete hyperelliptic integrals may be expressed in terms of generalised hypergeometric functions \cite{AbramowitzStegun1964}.

\subsection{Even number of branch points}

It is natural to inquire as to the existence of contour integrals of the type (\ref{H10a}) which may be regarded as the analogues of hyperelliptic integrals associated with an even number $2g+2$ of branch points. These hyperelliptic integrals arise in connection with the multi-phase averaged nonlinear Schr\"odinger (NLS) equations \cite{ForestLee1986}. In principle, the analogue of the solution (\ref{H6}) of the discrete Euler-Poisson-Darboux system, that is,
\bela{H24}
 \tilde{\varphi} = \prod_{k=0}^g p(\xi,n^{2k+1},0)\prod_{k=1}^{g+1}p(\xi,n^{2k},{\textstyle\frac{1}{2}}),
\ela
is still valid but it is seen that this ansatz does not lead to the distribution of poles and zeros in the case of odd ``genus'' by formally letting $n^{2g+2}\rightarrow\infty$. However, this situation may be rectified by annihilating the poles of $\tilde{\varphi}$ and introducing new poles in the ``non-singular'' regions by multiplication of $\tilde{\varphi}$ by an appropriate function of $\xi$ which has zeros and poles at half-integers and integers respectively. By virtue of the symmetries of the Gamma function, it turns out natural to introduce the ``complementary'' function 
\bela{H25}
  \varphi = -\tilde{\varphi}\cot\pi\xi.
\ela
Indeed, in terms of the ``complementary'' solution 
\bela{H26}
  q(\xi,n,\nu) = \frac{\Gamma(n + \nu -\frac{1}{2} - \xi)}{\sqrt{\delta}\,\Gamma(n + \nu - \xi)}
\ela
of the difference equation (\ref{E15a}) which is related to $p(\xi,n,\nu)$ by
\bela{H27}
  q(\xi,n,\nu) = p(\xi,n,\nu)\tan\pi(\xi-\nu),
\ela
it is readily verified that
\bela{H28}
 \varphi = q(\xi,n^{2g+2},{\textstyle\frac{1}{2}})\prod_{k=0}^g p(\xi,n^{2k+1},0)\prod_{k=1}^{g}p(\xi,n^{2k},{\textstyle\frac{1}{2}}).
\ela
Hence, the poles and zeros are distributed as required, that is, there are no zeros or poles in the intervals  $(n^{2\kappa-1}-1,n^{2\kappa})$. It is noted that, for convenience, the scaling of $q$ has been chosen in such a manner that it approximates the function $(x - \zeta)^{-1/2}$ rather than $(\zeta - x)^{-1/2}$ in the sense of (\ref{E15c}). Furthermore, up to a sign, $\varphi$ is symmetric in $p$ and $q$ due to the identity
\bela{H29}
  \textstyle p(\xi,n,0)p(\xi,m,\frac{1}{2}) = -q(\xi,n,0)q(\xi,m,\frac{1}{2})
\ela
for any integers $m$ and $n$. Once again, in the simplest case $g=1$, the contour integral (\ref{H11}), where the contour $b_1$ passes counterclockwise through the intervals $(n^1-1,n^2)$ and $(n^3-1,n^4)$, turns out to be a very good approximation of the corresponding elliptic integral
\bela{H30}
 \bear{c}\dis
  \phi^{\rm O} = \frac{2}{\pi\sqrt{(x^3-x^1)(x^4-x^2)}}\sfK\left(\sqrt{\frac{(x^3-x^2)(x^4-x^1)}{(x^3-x^1)(x^4-x^2)}}\right)\\[6mm]
  x^1 = \delta n^1,\quad x^2 = \delta(n^2 + \frac{1}{2}),\quad x^3 = \delta n^3,\quad x^4 = \delta(n^4 + \frac{1}{2})
 \ear
\ela
valid in the classical continuous case. In general, discrete $b$-cycles are defined as closed paths of counterclockwise orientation passing through the pairs of intervals \mbox{$(n^{2\kappa-1}-1,n^{2\kappa})$} and $(n^{2g+1}-1,n^{2g+2})$ for $\kappa = 1,\ldots,g$.

\section{Perspectives}

We conclude with a selection of open problems which naturally arise in connection with the theory presented in this paper. For instance, it has been pointed out in \cite{2,20,18,Dubrovin97} that the theory of semi-Hamiltonian systems of hydrodynamic type is closely related to the analysis of the critical points of appropriately chosen functions. In the current context, if $\psi$ is an eigenfunction satisfying the discrete conjugate net equations (\ref{E14}) and $\{\mu^i\}$,  $\{\lambda^i_\alpha\}$ are associated sets of adjoint eigenfunctions then one may introduce the corresponding Combescure transforms $\psi_\alpha$ and $\tilde{\Theta}$ according to
\bela{P1}
  \Delta_i\psi_\alpha = \lambda^i_\alpha\Delta_i\psi,\quad \Delta_i\tilde{\Theta} = \mu^i\Delta_i\psi.
\ela
The key function $\Theta$ is now defined by
\bela{P2}
  \Theta = x\psi + \sum_{\alpha=1}^Nt^\alpha\psi_\alpha - \tilde{\Theta},
\ela
where, {\em a priori}, $x$ and $t^\alpha$ are merely parameters. In analogy with the continuous case, critical points $\n_c$ of the function $\Theta$ are defined as points $\n$ where the ``discrete derivatives'' of $\Theta$ vanish, that is, ${\Delta_i\Theta|}_{\n=\n_c}=0$. Accordingly, we obtain
\bela{P3}
  x \Delta_i\psi(\n_c) + \sum_{\alpha=1}^N t^\alpha \Delta_i\psi_\alpha(\n_c) - \Delta_i\tilde{\Theta}(\n_c)=0
\ela
so that the definitions (\ref{P1}) imply that
\bela{P4}
 x + \sum_{\alpha=1}^Nt^\alpha\lambda^i_{\alpha}(\n_c) - \mu^i(\n_c) = 0.
\ela
The latter relate $x$ and $t^\alpha$ to $\n_c$ in the same manner (with the index on $\n_c$ being dropped) as the algebraic system (\ref{D22}) which gives rise to the discrete generalised hodograph equations (\ref{D24}). The implications of this observation are currently being investigated.

In the preceding, we have regarded ``complete'' hyperelliptic integrals as functions of their branch points $x^i$ and, in this context, put forward a canonical definition of their discrete analogues. It is natural to inquire as to the existence of similar analogues of ``incomplete'' hyperelliptic integrals and their associated differential equations. For instance, in the classical case, elliptic integrals are related by inversion to the differential equation
\bela{P5}
  \frac{d\zeta}{ds} = \sqrt{(\zeta - x^1)(\zeta - x^2)(\zeta - x^3)}
\ela
which essentially defines the elliptic Weierstrass $\wp$ function \cite{AbramowitzStegun1964}. It is evident that the approach pursued in this paper suggests that one should examine in detail the properties of the differential equation
\bela{P6}
\frac{d\zeta}{ds} = \delta^{\frac{3}{2}}
                \frac{\Gamma(\xi - n^1 + \frac{3}{2})}{\Gamma(\xi - n^1 + 1)}
                \frac{\Gamma(\xi - n^2 + 1)}{\Gamma(\xi - n^2 + \frac{1}{2})}
                \frac{\Gamma(\xi - n^3 + \frac{3}{2})}{\Gamma(\xi - n^3 + 1)},\quad \xi = \frac{\zeta(s)}{\delta}
\ela
which may be regarded as a one-parameter deformation of the classical differential equation (\ref{P5}). The latter is retrieved in the usual limit $\delta\rightarrow0$.

In \S 5, we have confined ourselves to a detailed discussion of the relevance of the discrete Euler-Poisson-Darboux-type system (\ref{E1}) for $\epsilon^i=\frac{1}{2}$. It is easy to see that, in the classical case, separable solutions of the Euler-Poisson-Darboux system (\ref{E1a}) for $\epsilon^i=\frac{1}{M}$, where $M$ is a positive integer, are obtained in terms of the superelliptic $(M,N)$-curves
\bela{P7}
  y^M = \prod_{i=1}^N(\zeta - x^i).
\ela
As in the hyperelliptic case $M=2$, the corresponding superelliptic integrals are relevant in the theory of Whitham-type equations. For instance, trigonal curves $(3,N)$ appear in connection with the Benney equations and the dispersionless Boussinesq hierarchy (see, e.g., \cite{GibbonsKodama94,BaldwinGibbons06} and references therein). It is therefore desirable to investigate whether it is possible to extend the theory developed in this paper to define canonical discrete analogues of superelliptic integrals and associated discrete characteristic speeds of Whitham type.

\section*{Acknowledgment}

B.G.K.\ acknowledges support by the PRIN 2010/2011 grant 2010JJ4KBA$\underline{\,\,\,}$003. W.K.S.\ expresses his gratitude to the DFG Collaborative Research Centre SFB/ TRR 109 {\em Discretization in Geometry and Dynamics} for its support and hospitality.


\begin{thebibliography}{99}

\bibitem{1} Courant, R,  Hilbert, D. 1989
{\sl Methods of mathematical physics}, vol.\ 2.
John Wiley \& Sons.

\bibitem{2} Whitham, GB. 1974
{\sl Linear and nonlinear waves}. 
John Wiley \& Sons.

\bibitem{3} Rozhdestvenskii, BL,  Yanenko, NN. 1980
{\sl Systems of quasilinear equations and their applications in gas dynamics}.
Transl.\ Math.\ Monographs, vol.\ 55.
Providence, RI: AMS.

\bibitem{4} Dubrovin, BA, Novikov, SP. 1983
Hamiltonian formalism of one-dimensional systems of hydrodynamic type and the Bogolyubov-Whitham averaging method. 
{\sl Soviet Math.\ Dokl.\ }{\bf  270}, 665--669.

\bibitem{5} Dubrovin, BA, Novikov, SP. 1984
On Poisson brackets of hydrodynamic type.
{\sl Soviet Math.\ Dokl.\ }{\bf 297}, 294--297.

\bibitem{6} Dubrovin, BA,  Novikov, SP. 1989
Hydrodynamics of weakly deformed soliton lattices. Differential geometry and Hamiltonian theory.
{\sl Russ.\ Math.\ Surveys} {\bf 44}, 35--124.

\bibitem{Per70} Peradzy\'nski, Z. 1970
On algebraic aspects of the generalized Riemann invariants method. 
{\sl Bull.\ Acad.\ Polon.\ Sci.\ S\'er.\ Sci.\ Tech.\ }{\bf 18}, 341--346.

\bibitem{FisPer76} Fiszdon, W, Peradzy\'nski, Z. 1976
Some geometric properties of a system of first-order non-linear partial differential equations.
In Fichera, G, ed. {\sl Trends in applications of pure mathematics to mechanics}. Monographs and Studies in Math., vol.\ 2. London: Pitman, 91--105.

\bibitem{Gru84} Grundland, AM. 1984 
Riemann invariants. 
In Rogers, C, Moodie, TB, eds. {\sl Wave phenomena: modern theory and applications}. North-Holland Math.\ 
Stud., vol.\ 97. Amsterdam: North-Holland, 123-152.

\bibitem{7} Tsarev, SP. 1985 
On Poisson brackets and one-dimensional systems of hydrodynamic type.
{\sl Soviet Math.\ Doklady} {\bf 31}, 488--491.

\bibitem{8} Tsarev, SP. 1991
The geometry of Hamiltonian systems of hydrodynamic type. The generalized hodograph method.
{\sl Math.\ in the USSR Izvestiya} {\bf 37}, 397--419.

\bibitem{9} Tsarev, SP. 1993
Classical differential geometry and integrability of
systems of hydrodynamic type. In {\sl Applications of analytic and geometric methods to nonlinear differential equations}. NATO ASI Series Volume 413, 241--249.

\bibitem{10} Darboux, G. 1910
{\sl Le\c{c}ons sur les syst\`emes orthogonaux et les coordonn\'ees curvilignes}.
Paris: Gauthier-Villars.

\bibitem{RogersSchief2002} Rogers, C, Schief, WK. 2002
{\sl B\"acklund and Darboux transformations. Geometry and modern applications in soliton theory}. 
Cambridge Texts in Applied Mathematics. Cambridge, UK: Cambridge University Press.

\bibitem{CourantFriedrichs1948} Courant, R, Friedrichs, KO. 1948
{\sl Supersonic Flow and Shock Waves}.
New York: Interscience Publishers Inc.

\bibitem{12} Darboux, G. 1887
{\sl Le\c{c}ons sur la th\'eorie g\'en\'erale des surfaces}, vol.1. 
Paris: Gauthier-Villars.

\bibitem{13} Kudashev, VR, Sharapov, SE. 1991
Inheritance of KdV symmetries under Whitham averaging and hydrodynamic symmetries of the Whitham equations.
{\sl Theor.\ Math.\ Phys.\ }{\bf 87}, 358--363.

\bibitem{14} Kudashev, VR, Sharapov, SE. 1991
Hydrodynamic symmetries for the Whitham equations for nonlinear Schr\"odinger equation (NSE).
{\sl Phys.\ Lett.\ A} {\bf 154}, 445--448.

\bibitem{15} Gurevich, AV, Krylov, AL, El, GA. 1991
Riemann wave breaking in dispersive hydrodynamics.
{\sl JETP Letters} {\bf 54}, 102--107.

\bibitem{16} Tian, FR. 1994
The Whitham-type equations and linear overdetermined systems of Euler-Poisson-Darboux type.
{\sl Duke Math.\ J.\ }{\bf 74}, 203--221.

\bibitem{17} Flaschka, H, Forest, MG, McLaughlin, DW. 1980
Multiphase averaging and the inverse spectral solution of the Korteweg-de Vries equation.
{\sl Commun.\ Pure Appl.\ Math.\ }{\bf 33}, 739--784.

\bibitem{18} Kodama, Y, Konopelchenko, B, Schief, WK. 
Lauricella functions, critical points and Whitham-type equations.
{\sl In preparation}.

\bibitem{19} Pavlov, MV. 2003
Integrable hydrodynamic chains. 
{\sl J.\ Math.\ Phys.\ }{\bf 44}, 4134--4156.

\bibitem{20} Konopelchenko, B, Martinez Alonso, L,  Medina, E. 2010
Hodograph solutions of the dispersionless coupled KdV hierarchies, critical points and the Euler-Poisson-Darboux equation.
{\sl J.\ Phys.\ A: Math.\ Theor.\ }{\bf 43}, 434020 (15pp).

\bibitem{21} Konopelchenko, B, Martinez Alonso, L,  Medina, E. 2013
Spectral curves in gauge/string dualities: integrability, singular sectors and regularization.
{\sl J.\ Phys.\ A: Math.\ Theor.\ }{\bf 46}, 225203 (27pp).

\bibitem{hyper} Belokolos, ED, Bobenko, AI, Enolskii, VZ, Its, AR, Matveev, VB. 1994
{\sl Algebro-geometric approach to nonlinear integrable equations}.
Springer Series in Nonlinear Dynamics.
Berlin: Springer-Verlag.

\bibitem{Eisenhart1962} Eisenhart, LP. 1962
{\sl Transformations of Surfaces}. 
New York: Chelsea.

\bibitem{Kudashev91} Kudashev, VR. 1991
Wave-number conservation and succession of symmetries during a Whitham averaging.
{\sl JETP Lett.\ }{\bf 54}, 175--179.

\bibitem{Ferapontov2000}  Ferapontov, EV. 2000
Systems of conservation laws within the framework of the projective theory of congruences: the L\'evy transformations of semi-Hamiltonian systems.
{\sl J.\ Phys.\ A: Math.\ Gen.\ } {\bf 33}, 6935--6952.

\bibitem{Suris03} Suris, YB. 2003
{\sl The problem of integrable discretization: Hamiltonian approach}.
Progress in Mathematics {\bf 219}.
Basel: Birkh\"auser.

\bibitem{AblowitzSegur1981} Ablowitz, MJ,  Segur, H. 1981
{\em Solitons and the inverse scattering transform}.
 Philadelphia: SIAM.

\bibitem{11} Bobenko, AI,  Suris, YB. 2009
{\sl Discrete Differential Geometry. Integrable Structure}.
Graduate Studies in Mathematics {\bf 98}.
Providence, RI: AMS.

\bibitem{KonopelchenkoBogdanov95}  Bogdanov, LV,  Konopelchenko, BG. 1995
Lattice and $q$-difference Darboux-Zakharov-Manakov systems via $\partial$-bar-dressing method.
{\sl J.\ Phys.\ A: Math.\ Gen.\ } {\bf 28}, L173--L178.

\bibitem{ManDolSan97} Ma\~nas, M, Doliwa, A, Santini, PM. 1997
Darboux transformations for multidimensional quadrilateral lattices. I.
{\sl Phys.\ Lett.\ A} {\bf 232}, 99--105.

\bibitem{LiuMan98} Liu, QP, Ma\~nas, M. 1998
Discrete Levy transformations and Casorati determinant solutions of quadrilateral lattices.
{\sl Phys.\ Lett.\ A} {\bf 239}, 159--166.

\bibitem{AkhmetshinKricheverVolvoski99} Akhmetshin, AA, Krichever, IM, Volvoski, YS. 1999
Discrete analogs of the Darboux-Egoroff metrics.
{\sl Proc.\ Steklov Institute of Mathematics} {\bf 225}, 16--39.

\bibitem{KonopelchenkoSchief1993} Konopelchenko, BG, Schief, WK. 1993
Lam\'e and Zakharov-Manakov systems: Combescure, Darboux and B\"acklund transformations.
{\sl Preprint AM93/9 Department of Applied Mathematics}, The University of New South Wales.

\bibitem{Hirota2003} Hirota, R. 2003 
How to obtain $N$-soliton solutions from $2$-soliton solutions.
{\sl RIMS  K\^{o}ky\^{u}roku} {\bf 1302}, 220--242.

\bibitem{DCN} Konopelchenko, BG,  Schief, WK. 1998
Three-dimensional integrable lattices in Euclidean spaces: conjugacy and orthogonality.
{\sl Proc.\ R.\ Soc.\ London A} {\bf 454}, 3075--3104.

\bibitem{DoliwaSantini97} Doliwa, A,  Santini, P. 1997
Multidimensional quadrilateral lattices are integrable.
{\sl Phys.\ Lett.\ }{\bf 233}, 265--272.

\bibitem{Ferapontov02} Ferapontov, EV. 2002
Invariant description of solutions of hydrodynamic-type systems in hodograph space: hydrodynamic surfaces.
{\sl J.\ Phys.\ A: Math.\ Gen.\ }{\bf 35}, 6883--6892.

\bibitem{IDT} Matveev, VB, Salle, MA. 1991
{\em Darboux transformations and solitons}.
Berlin Heidelberg: Springer-Verlag.

\bibitem{AbramowitzStegun1964} Abramowitz, M,  Stegun, IA, eds. 1964
{\sl Handbook of mathematical functions with formulas, graphs, and mathematical tables}.
New York: Dover Publications; {\sl NIST Digital Library of Mathematical Functions}, http://dlmf.nist.gov/

\bibitem{TricomiErdelyi1951} Tricomi, F, Erd\'elyi, A. 1951
The asymptotic expansion of a ratio of Gamma functions.
{\sl Pacific J.\ Math.\ }{\bf 1}, 133--142.

\bibitem{GelfandGraevRetakh92} Gelfand, IM, Graev, MI, Retakh, VS. 1992
General hypergeometric systems of equations and series of hypergeometric type.
{\sl Russ.\ Math.\ Surveys} {\bf 47}, 1--88.

\bibitem{ForestLee1986} Forest, MG,  Lee, J-E. 1986
Geometry and modulation theory for periodic nonlinear  Schr\"odinger equation. In Dafermos, C {\em et al.}, eds.
{\sl Oscillation theory, computation and methods of compensated compactness}. 
IMA Volumes on Mathematics and Its Applications {\bf 2}. New York: Springer-Verlag, 35--69.

\bibitem{Dubrovin97} Dubrovin, B. 1997
Functionals of Peierls-Fr\"ohlich type and variational principle for Whitham equations.
{\sl Amer.\ Math.\ Soc.\ Transl.\ }{\bf 179}, 35--44.

\bibitem{GibbonsKodama94} Gibbons, J, Kodama, Y. 1994
Solving dispersionless Lax equations. In Ercolani, NM {\em et al.}, eds.
{\sl Singular limits of dispersive waves}. New York: Plenum Press, 61--66.

\bibitem{BaldwinGibbons06} Baldwin, S, Gibbons, J. 2006
Genus 4 trigonal reduction of the Benney equations.
{\sl J.\ Phys.\ A: Math.\ Gen.\ }{\bf 39}, 3607--3639.

\end{thebibliography}
\end{document}